# PRELIMINARY DESIGN OF MUSES CONTROL SYSTEM BASED ON RT-CORBA AND JAVA

Toshiya Tanabe, Toshikatsu Masuoka, Jun-ichi Ohnishi, Motonobu Takano
and Takeshi Katayama, RIKEN, 2-1 Hirosawa, Wako, Saitama 351-0198, JAPAN

Abstract

Common Object Request Broker Architecture (CORBA) based control system has been utilized for the first phase of RIKEN-RI Beam Factory (RIBF) [1] at the developing stage. Software sharing with Jozef Stefan Institute (JSI) in Slovenia via CORBA/JavaBeans has been successfully demonstrated. Accelerator Beans (Abeans) [2] components developed in JDK1.2.2 have been ported to RIKEN's CORBA server. The second phase of the RIBF project is named "Multi-Use Experimental Storage rings" (MUSES) project, which includes an accumulator cooler ring (ACR) and collider rings. Due to the much larger number of controlled objects and more stringent timing requirement than the first-phase project, we contemplate using recently established real-time (RT) CORBA specification [3]. Summary of our efforts to test RT-CORBA with the existing JaveBeans components and other related subjects are described in this paper.

## 1 INTRODUCTION

CORBA is the standard distributed object architecture for an open software bus, which allows object components created by different OSs and languages to interoperate. It would make integration with legacy controls easier, which is suited for RIBF project as it is an expansion of existing facilities. Java language [4] is known to be very well thought-out objected oriented language and its most useful feature is platform independence using interpreter and Java Virtual Machine (VM) which executes the byte code.

In our previous R&D efforts [5], we have confirmed interoperability through IIOP among different ORBs, especially ones for board computers such as VME and CompactPCI. It has also been shown that the performance degradation compared to pure socket communication is insignificant.

Recent Introduction of RT-CORBA specification allows us to utilize the full capability of RTOS such as VxWorks [6]. RT-CORBA augments the shortcomings of regular CORBA 2.x specs, namely the lack of quality of service (QoS) controls, real-time programming features and performance optimizations. Among various features of RT-CORBA, we have examined mainly processor resources control such as thread pools, priority mechanism and intra-process mutexes. VisiBroker for Tornado 4.1a and ORBexpress-RT 2.3.2 [7] were chosen for comparison.

Software sharing with Jozef Stefan Institute (JSI) which allows us to use their GUIs by JaveBeans for power supply controls in our Intermediate-stage Ring Cyclotron (IRC). Another reason which encourages us to use CORBA is the wide use of Programmable Logic Controls (PLCs) in Japanese manufacturers. Each company has its favorite type of PLCs that are often used to construct a closed system such as RF cavity controls and vacuum system. CORBA-Experimental Physics and Industrial Control System (EPICS) server have been developed to make a seamless connection with our legacy controls upstream. Some issues on Mathematica® -Java integration will also be discussed.

## 2 REAL-TIME CORBA TESTS

### 2.1 Mutex

RT-CORBA mutex operations would conceal the difference in thread-synchronizing mechanism in the underlying OS. We have compared the durations of time when *lock()/unlock()* operations are repeated for 1000 times for a) VxWorks native semTake(), semGive(), b) rtMutex -> lock(), rtMutex -> unlock() for VisiBroker, c) the same for ORBexpress. The elapsed time was measured by "WindView" tool [6]. The results (average time for 1000 operations in msec and standard deviation) are shown in Table 1. It appears that extra latency caused by RT-CORBA abstraction is less than 1 μsec and considered to be negligible for most accelerator applications.

Table 1: Comparison between VxWorks semaphore and rtMutex

|  | Semaphore | VisiBroker | ORBexpress |
|---|---|---|---|
| Ave. (ms) | 2.092 | 2.334 | 2.742 |
| St. Dev. | 0.00625 | 0.00833 | 0.005 |

### 2.2 Thread Pools and CORBA Priority

Thread pools can be defined and associated with POAs in an RT-CORBA server. They can have both a fixed number of static threads, and dynamic threads whose maximum number should be specified.

With a proper priority mapping, RT-CORBA priorities can also conceal the variations of native priority mechanism of RTOSs. For example, VxWorks has 0 for highest priority and pSOSystem has 255 for the highest. RT-CORBA priority always has 32767 for the highest. There are two types of PriorityModel policy, "server declared" and "client propagated". We have set ORB's priority 25 and that of VxWorks 155, and measured the connection time for both server-declared model and client-propagated model. For both static and dynamic threads, two threads can be created in the test program. Since ORBexpress-RT 2.3.2 lacks *create_threadpool()* API, this test was conducted only in VisiBroker.

First, with the help of WindView, it was confirmed that VxWorks tasks could be switched based on CORBA priorities. Table 2 show the time required to establish the connection between a client and a servant for the server-declared model. For dynamic threads, the default timeout to destroy them is 300 seconds. Measurement was done both cases, one with existing threads and one after their disappearance. The same type of measurement was done for client-propagated model and the results are shown in Table 3.

Table 2: Connection time for server declared model

| **Static Thread** | | |
|---|---|---|
| | Init. Connection | From 2$^{nd}$ Connection |
| Ave (ms) | 4.623 | 3.308 |
| St. Dev. | 0.008 | 0.143 |
| **Dynamic Thread** | | |
| | Init. Conn. | From 2$^{nd}$ w. threads | From 2$^{nd}$ w.o. threads |
| Ave (ms) | 4.800 | 3.326 | 3.581 |
| St. Dev. | 0.012 | 0.125 | 0.152 |

Table 3: Connection time for client propagated model

| **Static Thread** | | |
|---|---|---|
| | Init. Connection | From 2$^{nd}$ Connection |
| Ave (ms) | 4.621 | 3.491 |
| St. Dev. | 0.016 | 0.175 |
| **Dynamic Thread** | | |
| | Init. Conn. | From 2$^{nd}$ w. threads | From 2$^{nd}$ w.o. threads |
| Ave (ms) | 4.801 | 3.545 | 3.801 |
| St. Dev. | 0.010 | 0.190 | 0.195 |

The results indicate that the initial connection always takes longer time as POA manager has to be activated. However, from the second time and on, there is not much difference among different cases we tested.

## 3 SOFTWARE SHARING WITH JSI

The possibility of software sharing for the accelerator controls has been discussed for a long time. However very few examples with the exception of EPICS and cdev [8] are found in reality. One of the reasons for the failure is that there is no standardization of data structure and communication protocol. Recent advance in XML for standard data format and CORBA/IIOP for standard inter-object communication may lead to more sharing in the future. We are currently working on the use of XML files for equipment parameter files to be stored in a database and parsed in VME. As for the GUIs, JavaBeans is one of the most promising candidates for sharable software. ANKA control system utilize this technology for their GUI and control logics as well as CORBA for communication. People at JSI where this set of software was developed kindly allow us to examine the possibility of sharing codes in our system.

Even though JSI people provide Accelerator Corba Interface (ACI) for the connection to their Accelerator Beans (abeans) which represent control logic, it is not necessarily easy to adapt ACI for our local system. That's partly because some of our hardware have extra functionalities that are not covered by ACI. Hence writing CORBA wrapper program with our Interface Definition Language (IDL) was chosen for convenience. Establishing agreeable IDLs for accelerator controls seems to be the most difficult task for the software sharing using CORBA.

The magnetic field measurement for sector magnets of intermediate-stage ring cyclotron (IRC) has been conducted using CORBA based system for some time. One large power supply (PS) is used for the main magnet per sector and 30 middle-sized PSs are used for the trim coils. Our PS object model is used to control all the PSs through VME master-slave type field bus (Hitachi-Zosen NIO-C & NIO-S [9]). The master board and the slave board for the main magnet PS, and that for the first trim coil PS are connected by optical fibers. The rest of the trim coil PSs are connected by IEEE485 in series.

## 4 PLC WRAPPER OBJECT

PLCs are Japanese manufacturer's favorite control tools for factory automation. There are many different kinds and they are practically incompatible among each other. Therefore one needs a mechanism which can conceal these differences for upper-level software. We have developed CORBA PLC wrapper programs for the PLC by OMRON [SYSMAC CVM1/CV] and they are used for RF control, vacuum monitoring, and motor-drive controls. PLC CORBA object in VME has an interface with Abeans, and read/write the data in the memory area in the PLC through UDP using FINS command that is provided by the manufacturer.

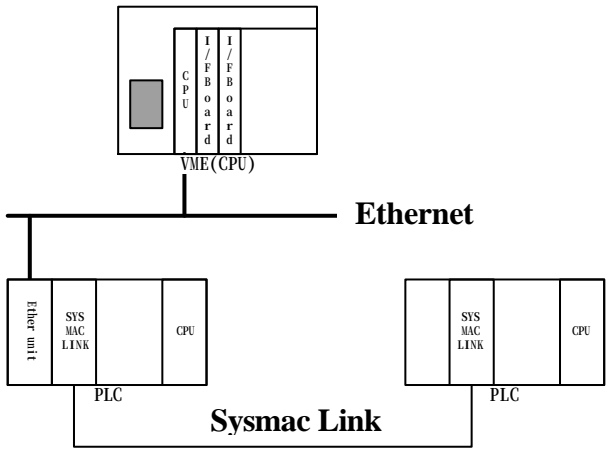

Fig. 1: PLC Wrapper system configuration.

## 5  CORBA-EPICS SERVER

RARF controls are currently being upgraded with EPICS. Hence a mechanism must be developed to allow for the old and the new controls to communicate with each other. CORBA is often used to integrate legacy systems with various kinds of wrapper software. However, in our case, it is easier to set up a gateway computer that acts as a translator between the two systems. The CORBA-EPICS (CE) server scheme fulfils this need.

The CORBA server used for the test is the one used for the main-coil power supply of an IRC. The version of EPICS installed in a Sun Workstation (Ultra10 Model 360, Solaris 2.6) was R3.13.2 and a portable channel access (CA) server program was running in it. The ORB is VisiBroker for C++ ver. 3.3. Sun Workshop3 C++ 4.2 was used for the compiler. Since CORBA and EPICS environments are in different network segments, some routing operation was required. Figure 2 shows the system configuration for this server.

The operation from EPICS to CORBA is initiated by a CA client. It specifies the records to process in the CE server and a modified portable CA server program passes the data to the CORBA client program. This CORBA client connects CORBA servers in the CORBA domain. A sample program (excas) has been modified to accommodate the following functions in the CA server:

From CORBA to EPICS, the CORBA client has to know the records to be accessed in the CA server in the EPICS domain. After the CORBA server program receives the call from the CORBA client, it calls a CA client in the CE server to connect to CA servers in the EPICS domain.

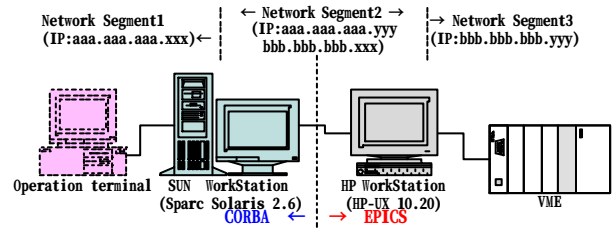

Fig. 2: CORBA-EPICS server system configuration.

## 6  MATHEMATICA®-JAVA LINK

Physicists would always like to use tools they are familiar with in their office. Mathematica is one of the most powerful tools for analysis in Physics. Wolfram research which produces Mathematica has recently distributed a free toolkit called JLink® which allows one to call Java from Mathematica and to use its kernel from a Java program.

When some of Abeans' GUI programs are called from Mathematica, one should distinguish whether the class called is the subclass of *java.awt.Window* or not. If not, the JFrame class has to be provided to draw it. From Java to Mathematica, one must be aware that Mathematica kernel is not multi-threaded.

## 7  CONCLUSION

Under the current circumstances, no vendor can be trusted through a lifetime of an accelerator. Therefore, a modern control system ought to be vendor independent as much as possible. We have demonstrated that RT-CORBA provides good abstraction to RTOS environment. JavaBeans for upper-level applications which communicate with front-end computers through RT-CORBA would result in a control system that could withstand ever more whimsical vendors' discontinued product lines.